 \documentclass{article}
\usepackage[dvips]{graphicx}
\usepackage{epsfig}

\title{Rounding of aggregates of biological cells: Experiments and simulations}

\author{Jos\'e C.M. Mombach$^1$, Damien Robert$^2$, Fran\c{c}ois
Graner$^3$, \\Germain Gillet$^4$, Gilberto L. Thomas$^5$, Marco
Idiart$^5$,\\ Jean-Paul Rieu$^2$\\
\\ (1) Laborat\'orio de Bioinform\'atica e Biologia Computacional \\
Universidade do Vale do Rio dos Sinos, Av. Unisinos, 950
\\ 93022-000 S\~ao Leopoldo, RS, Brazil. \\
(2) Laboratoire de Physique de la Mati\`ere Condens\'ee et\\
Nanostructures, UMR5586 CNRS and University Lyon 1, \\43 boulevard
du 11 Novembre, 69623 Villeurbanne, France. \\(3) Laboratoire de
Spectrometrie Physique,\\ UMR5588 CNRS and Universit\'e Grenoble I
BP 87, \\38402 St. Martin D'Heres Cedex, France.
\\ (4) Institut de Biologie et Chimie des Prot\'eines, CNRS and University \\Lyon 1,
7 passage du Vercors, 69367 Lyon Cedex 07, France.\\(5) Instituto
de F\'{\i}sica, Universidade Federal do Rio Grande do Sul,\\Av.
Bento Gon\c{c}alves, 9500 91501-970 Porto Alegre, RS, Brazil.}
\date{}
\begin{document}

\maketitle

\begin{abstract}
The influence of surface tension and size on rounding of cell
aggregates are studied using chick embryonic cells and numerical
simulations based on the cellular Potts model. Our results show
exponential relaxation in both cases as verified in previous
studies using 2D Hydra cell aggregates. The relaxation time
decreases with higher surface tension as expected from
hydrodynamics laws. However, it increases faster than linearly
with aggregate size. The results provide an additional support to
the validity of the cellular Potts model for non-equilibrium
situations and indicate that aggregate shape relaxation is not
governed by the hydrodynamics of viscous liquids.

\end{abstract}

\maketitle

\section{Introduction}

It is  now well established that certain living tissues behave as
viscoelastic fluids \cite{Beysens}. Classical experiments on the
relaxation of embryonic aggregates subjected to mechanical
deformations, either by centrifugation \cite{Phillips} or by
compression \cite{Forgacs}, demonstrated that tissues relax as
elastic materials on short time scales and as viscous liquids on
long time scales. The relaxation of compressed cellular aggregates
was analyzed using a Kelvin model of viscoelasticity
\cite{Forgacs}. As stated by the authors, this model may represent
a strong oversimplification of the tissue rheology but it gives an
estimation of the tissue elastic and viscous parameters
\cite{Forgacs}. The fusion or the rounding up of cell aggregates
are similar to the coalescence or rounding up of liquid droplets.
Gordon {\it et al.} showed that these processes are driven by
tissue surface tensions $\sigma$ and opposed  by tissue viscosity
$\eta$ making a formal analogy with the rheology of viscous
liquids. They obtained first estimates of the ratio $\eta / \sigma$
for chicken embryonic cells \cite{Gordon}. However, the authors
gave only orders of magnitude of this ratio as they performed only
preliminary experiments, presumably because it was difficult to
obtain suspended three dimensional aggregates which were fixed
over long times (rotation or flattening on a surface are
undesirable). Later, the surface tensions of different chicken
embryonic tissues have been measured using a parallel plate
compression apparatus \cite{Foty, Foty2}.

In previous investigations one of us showed that a two dimensional
(2D) assembly of Hydra cells behaves like a fluid: (i) cells
exhibit collective \cite{Rieu, Rieu3} or random \cite{Rieu2}
motion, (ii) the hydrodynamical laws also apply for the
rounding-up of initially elliptical  aggregates \cite{Rieu3}. The
shape relaxation was analyzed using the 2D hydrodynamical
equations derived by Mann {\it et al.} for polymer monolayers in
the case of dissipation dominated by surface viscosity
\cite{Mann}. This method allowed an estimation
$\eta/\sigma\simeq0.5$ min/$\mu$m although $\sigma$ was not
measured independently. Compared to the relaxation analysis of
free 3D aggregates, 2D aggregates offer some advantages: they are
not moving, nor rotating or flattening on a nearby surface. The 2D
system also allows to track simultaneously cell motion and a
direct comparison with 2D simulations.

In this paper, we summarize the main results obtained previously
on 2D Hydra aggregates \cite{Rieu3}. We also present current
experiments on the rounding of 3D aggregates of chicken embryonic
cells (neural cells from the retina) and simulations of rounding
of 2D aggregates using the cellular Potts model. Because of its
flexibility and simplicity of implementation, Potts model has
become a common technique for cell level simulation of biological
tissues. It describes effectively the global features of tissue
rearrangement experiments, including cell sorting \cite{glazier}
and even the cell movements during the entire life cycle of the
slime mould {\it Dictyostelium discoideum} \cite{Maree}. In this
study, we show that Potts model effectively captures the
exponential relaxation dynamics of aggregate rounding. However,
both new experiments and simulations show a deviation from the
hydrodynamics of viscous liquids.

\section{Results}

\subsection{Rounding of 2D Hydra aggregates}

Hydra aggregates were prepared according to previously reported
methods \cite{Rieu} with 25 $\mu m$ spacers (Fig. 1A). A sequence
of events in a pure endodermal aggregate initially elliptical is
displayed in Figs. 1B-C. Rounding up is completed after 6h. We
studied also the rounding of mixed and pure ectodermal aggregates
\cite{Rieu3}. Images were digitized at intervals of $5\sim 10$ min
and analyzed on a computer using the software NIH Image
\cite{NIH}. The deformation parameter $d=\frac{M}{m}-1$ was used
to describe the aggregate shape, where $M$ and $m$ are the major
and minor axis of the fitted ellipse. $d$ is zero for a perfect
circle and increases for elongated aggregates.

Fig. 1D shows a log-linear plot of the time dependence of the $d$
for the different 2D aggregates investigated. For each aggregate,
$d$ decreases exponentially with time. Such exponential behavior
is expected theoretically in 2D from the hydrodynamics laws in
case of viscosity dominated by the surface viscosity and rounding
driven by line tension \cite{Mann}. In case of 2D aggregates
confined between two glass slides, if the cell/glass friction may
be neglected compared to cell/cell friction, the relaxation time
is given by relation:
\begin{equation}
 T_c = \frac{\eta}{\sigma} R\;,
\label{tc}
\end{equation}
where $R$ is the radius of the rounded aggregate (final time),
$\eta$ is a bulk surface viscosity, and $\sigma$ the
tissue/external medium surface tension \cite{Rieu3}. The typical
time scale of rounding depends thus on the ratio between the
surface tension and viscosity. The aggregates round to decrease
superficial tension, while viscosity acts in the sense of delaying
the rounding. Fig. 1E shows that for each aggregate type, $T_c$ is
increasing with $R$ as expected from the hydrodynamical law Eq.
\ref{tc}. The ratio $\eta / \sigma = 0.5 \pm 0.2$ min/$\mu$m which
gives $\eta=3 \times 10^4$ Pa.s (1 Pa.s = 10 Poise) using
$\sigma$=1 mN/m as a typical value of tissue surface tension
\cite{Forgacs, Foty, Foty2}. This is $10^{7}$ times higher than
the water viscosity.

\subsection{Rounding of 3D chicken embryonic neural retina aggregates}

We performed relaxation experiments on chicken embryonic tissues
which have been already characterized by surface tensiometry
\cite{Forgacs,Foty}. Neural cell tissues were obtained from the
retina of 9 day old chicken embryos. After dissection and
trypsinization, cells were reaggregated by shaking them slowly in
12-well plates (200 r.p.m.) containing culture medium (BME culture
medium with 10\% fetal Calf Serum \cite{Gillet}) at 37$^\circ$C
during 16 hours. We could not obtain proper 2D chicken aggregates
for relaxation experiments. We used instead non-compressed 3D
aggregates, just floating near a bottom slide glass (Fig. 2A).
These aggregates are sufficiently stable for shape analysis, they
do not move or rotate significantly and do not spread on the
bottom slide glass. Figures 2B-C show the shape of a 9-day old 3D
aggregate between initial and final time (24h). The time
dependence of the deformation parameter of these aggregates is
well exponential as expected from hydrodynamics  with a relaxation
time $T_c$ of about 1 day (Fig. 2D). In case of rounding of 3D
ellipsoid droplets, $T_c$ is given by a slightly different
relation than the 2D equation \ref{tc} \cite{Gordon}:
\begin{equation}
 T_c = \beta \frac{\eta}{\sigma} R\;,
\label{tc3D}
\end{equation}
where the constant $\beta \sim 0.95$ in case of aggregate internal
viscosity much larger than external medium viscosity
\cite{Gordon}, and $R$ is the radius of the rounded 3D aggregate.

At a first sight, $T_c$ seems to increase linearly with aggregate
radius. The doted line in Fig. 2E corresponds to such a linear fit
with Eq. \ref{tc3D} (correlation coefficient, 0.880). The ratio
$\eta / \sigma$ is $16.8 \pm 4.5$ min/$\mu$m. The surface tension
of neural aggregates dissected at 6 days and maintained in culture
for various time periods was measured with the compression plate
apparatus \cite{Foty2}. The value increases with continued
incubation for about a day (between 6 and 7 days), then stabilizes
for a time period of about 2 days at $\sigma=1.6 \pm 0.6$ mN/m and
increases again between 8 and 9 days to about $\sigma=4.0 \pm 1.0$
mN/m. Taking this latter value, we obtain the following neural
tissue viscosity value at 9 days: $\eta \simeq (4 \pm 2) \times
10^6$ Pa.s. It is much higher than in Hydra aggregates. This is
not surprising. First, because the kinetics of rounding is much
slower in chicken embryonic neural aggregates. Also because cell
packing is much higher in these 3D aggregates (Fig. 2B-C) than in
the 2D Hydra aggregates (Fig. 1B-C). It is likely that tissue
surface tension and tissue viscosity depend both on cell/cell
adhesion \cite{Rieu3} and that adhesion is higher for 3D chicken
embryonic aggregates than 2D Hydra aggregate.

However, $T_c$ is probably increasing faster than linearly, {\it
i.e.}, a power law $\sim R^{1.52}$ fits better the data
(correlation coefficient, 0.939). Although there is large
dispersion of the experimental points which is often inherent to
such a biological system, the visual impression is also better
than with linear fit (Fig. 2E). Of course, these data should be
confirmed but they seem to indicate that 3D chick embryonic cell
aggregates do not behave as viscous liquids. The relaxation of
initially elongated aggregates toward a circular shape is clearly
exponential but relaxation time is not proportional to the
aggregate size.

\subsection{Simulations}

The model is based on the cellular Potts model \cite{glazier} and
is defined as follows: at each site $(i, j)$ of a square lattice
with dimensions $L\times L$ we attribute a spin that may assume
any integer value. The set of all sites with equal spin $S$
defines a cell, labeled by S.

Adhesion between cells originates in the interaction between each
spin and its 20 surrounding neighbors (up to fourth neighbors to
avoid pinning to the square lattice \cite{holm}). The lowest
interaction energy, here taken as zero, happens between equal
spins, simulating the absence of surface tension between sites
belonging to the same cell. Energy between different cells is
taken to be positive and when different types of cells are
considered, the interaction energy may also be different,
depending on the types involved in the interaction. The intensity
of each possible interaction must be specified. Hence, for
example, for a system made of an aggregate of cells of two types
only and medium we must define three parameters. The result is a
proportional to the shared interface length, with a
proportionality constant depending on the neighboring cell types
to reflect differential adhesion strengths.

Cell resistance to compression is modeled by an energy term
proportional to the square deviation of cell area from its target
area $A_T (S)$. The proportionality parameter $\lambda$ plays the
role of a Lagrange multiplier for this constraint and regulates
cell compressibility. The complete energy used in the Monte Carlo
protocol is
\begin{equation}
H=\sum_{ij}\sum_{i^{\prime}j^{\prime}}E_{S_{ij}S_{i^{\prime}j^{\prime}}}
\left(1-\delta_{S_{ij},S_{i^{\prime}j^{\prime}}}\right)+
\lambda\sum_S\left[a(S)-A_T(S)\right]^2\;, \label{hamil}
\end{equation}
where $a(S)$ is the area of cell $S$, $S_{ij}$ is the spin at site
$(i,j)$, and $E_{S_{ij},S_{i^{\prime}j^{\prime}}}$ is the
interaction energy between neighboring sites labeled by $S_{ij}$
and $S_{i^{\prime}j^{\prime}}$. If
$S_{ij}=S_{i^{\prime}j^{\prime}}$ then
$E_{S_{ij},S_{i^{\prime}j^{\prime}}}=0$ since in this case both
sites belong to the same cell.

The aggregate surface tension ($\sigma$, at $T$=0) derives from
the coupling energies in Eq. (\ref{hamil}) by \cite{glazier}
\begin{equation}
\sigma=E_{cM}-\frac{E_{cc}}{2}\;, \label{st}
\end{equation}
where $E_{cM}$ and $E_{cc}$ are the adhesion energies per unit of
interface between cells and medium and between cells,
respectively.

We implement the simulations as follows. We
randomly choose a lattice site $(i,j)$ and one of its eight
surrounding neighbors. We then propose that the site assumes the
neighboring spin value $S^{\prime}$ and accept with probability
$P$:

$$
P(S\rightarrow S^{\prime})=
    \left\{ \begin{array}{ll} \exp(-\Delta
H/T)
   & \mbox{if $\Delta H>0$}\;, \\
\\
     1
   & \mbox{if $\Delta H\leq 0$}\;,
                       \end{array} \right.
$$
where $\Delta H$ is the variation in energy produced by the change
of spin values. A Monte Carlo Step (MCS), is
defined as $L^2$ exchange attempts, the total number of sites in
the lattice. For sake of simplicity, we use the MCS as the unit of ``time"
in what follows. In fact, the random sampling of spins, and the immediate dissipation of their energy (perfectly local, both in time and space) imply that the correspondence with real time
is not straightforward \cite{binder}. 

In this model the simulation temperature $T$ relates to active
cell membrane motion. Higher $T$ implies higher membrane activity
and vice-versa. $T$ simulates membrane fluctuations driven by the
cytoskeleton in real cells and should not be confused with much
smaller thermal fluctuations \cite{glazier}.

Aggregate rounding was investigated using as initial states
elliptical aggregates with different numbers of cells: 544, 953,
1642, 2204, and 3233 cells. Cells areas have approximately 100
lattice sites and cell-medium surface tensions have values: 4.5,
5.5, 6.5, 7.5, 8.5, 9.5. To generate these surface tensions we
keep constant the interaction energy between cells, $E_{cc}$=1,
and vary the energy between cells and medium, $E_{cM}=$ 5, 6, 7,
8, 9, and 10. The simulation temperature used is 10 and
$\lambda=1$. For aggregates with 1642 cells we simulated all of the above
surface tensions, however for other aggregate sizes we
studied only the three lower values of surface tension. In Figs.
3A and 3B we show, respectively, the initial and final state
(complete rounding-up) of the simulated aggregate with 544 cells.

We calculate the deformation in the simulated aggregates using the
ratio of the eigenvalues of the two dimensional variance of mass
distribution of the aggregate that is defined as
\begin{equation}
\left(
\begin{array}{cc}
  \langle x^2\rangle-\langle x\rangle^2 & \langle
  xy\rangle-\langle x\rangle\langle y\rangle\\
  \langle xy\rangle-\langle x\rangle\langle y\rangle &
  \langle y^2\rangle-\langle y\rangle^2 \\
\end{array}
\right)
\end{equation}
where $x$ and $y$ are the coordinates of all lattice sites
belonging to cells in the aggregate. As the aggregate rounds, the
ratio between the two eigenvalues of the matrix above, $e_1$ and
$e_2$ goes to 1. In analogy with experimental studies, we define
aggregate deformation using the ratio between the larger and lower
eigenvalues of the matrix minus one: $d=e_1/e_2-1$, $(e_1>e_2)$.
This measure of the deformation differs from that  used in the
experiments but should yield the same exponential decay.

In Fig. 4 we plot $d$ as function of time for aggregates with 1642
cells for surface tensions 4.5, 6.5, 8.5, and 9.5. The relaxation
is clearly exponential in all cases. Fig. 5 shows $T_c$ as a
function of $\sigma^{-1}$ for simulated aggregates with 1642 cells
and surface tensions 4.5, 5.5, 6.5, 7.5, 8.5, and 9.5. A
statistical analysis of the fit in the figure supports an
intercept of the line with the vertical axis at the origin
(p-value of linear coefficient = 0.8), consistent with Eq.
\ref{tc}. According to Eq. \ref{tc} the slope of the fit is the
R$\times$viscosity, however simulations predict an exponent higher
than 1 from a log-log plot of $T_c$ vs. aggregate radius. As we
can see in Fig. 6, we determined that the slope for simulation is
about 2.5 and for chicken data is about 1.5 (see also fit in Fig.
2). Simulation exponent seems robust to variations in temperature
and surface tension. In both cases the results suggest a deviation
from hydrodynamics.

\section{Conclusions}
We have presented new experimental results on the relaxation of
biological cell aggregates (chick embryonic neural retina
aggregates). Our motivation was to measure the equivalent
viscosity of this tissue. Such measurements are important to
characterize cell adhesiveness and to provide model parameters for
theoretical or numerical models of collective motion
\cite{Palsson}. The relaxation of 3D aggregates is exponential,
however, contrary to 2D aggregates of Hydra cells \cite{Rieu3},
relaxation time seems to increase faster than linearly with
aggregate size. We are currently working on the verification of
this experimental dependence in case of 3D aggregates. To this end, 
we need to improve the experimental environment (C0$_2$ and 0$_2$
control) in order to increase the observation time to a few days
and be able to study larger aggregates. It should be also noted
that 2D results suffer large errors. However if this dependence is
confirmed, the reason why 2D Hydra aggregates behave as viscous
fluid but not the 3D chick aggregates is not completely clear. The
answer may arise from cell trajectory studies. In case of 2D hydra
aggregates, all cells participated to rounding, motion was very
collective \cite{Rieu3}. We have not performed a similar study yet
but it seems that in larger 3D aggregates, intense cell movements
occur mainly in the aggregate periphery in a band of a few cell
diameter while in the core of the aggregate cells are less active.
Assuming nevertheless a fluid behavior, we obtain the following
estimate for the tissue viscosity of 9 day old neural retina,
$\eta \simeq (4 \pm 2) \times 10^6$ Pa.s.

A simple Potts model simulation of the phenomena captures the
exponential decay and indicates a superlinear dependence of
relaxation time with aggregate size as observed in experiments
with chicken cells. The simulation exponent, however does not
agree with experiments but a possible explanation for the
difference may come from cell trajectory studies. 3D simulations might also 
provide useful quantitative comparisons with 3D aggregates.

\section*{Acknowledgments}
We acknowledge useful discussions with Rita M. C. de Almeida, and
M. Aubouy. This work is a brazilian-french collaboration supported
by Capes and Cofecub agencies project n$^\circ$ 414/03. J.C.M.
Mombach acknowledges the partial support of HP Brazil R\&D. J.P.
Rieu and G. Gillet acknowledge the partial support of french
C.N.R.S. (ACI Dynamique et R\'eactivit\'e des Assemblages
Biologiques).


\begin{figure}[ht]
\centerline{
\epsfxsize 10truecm
\epsfbox%
{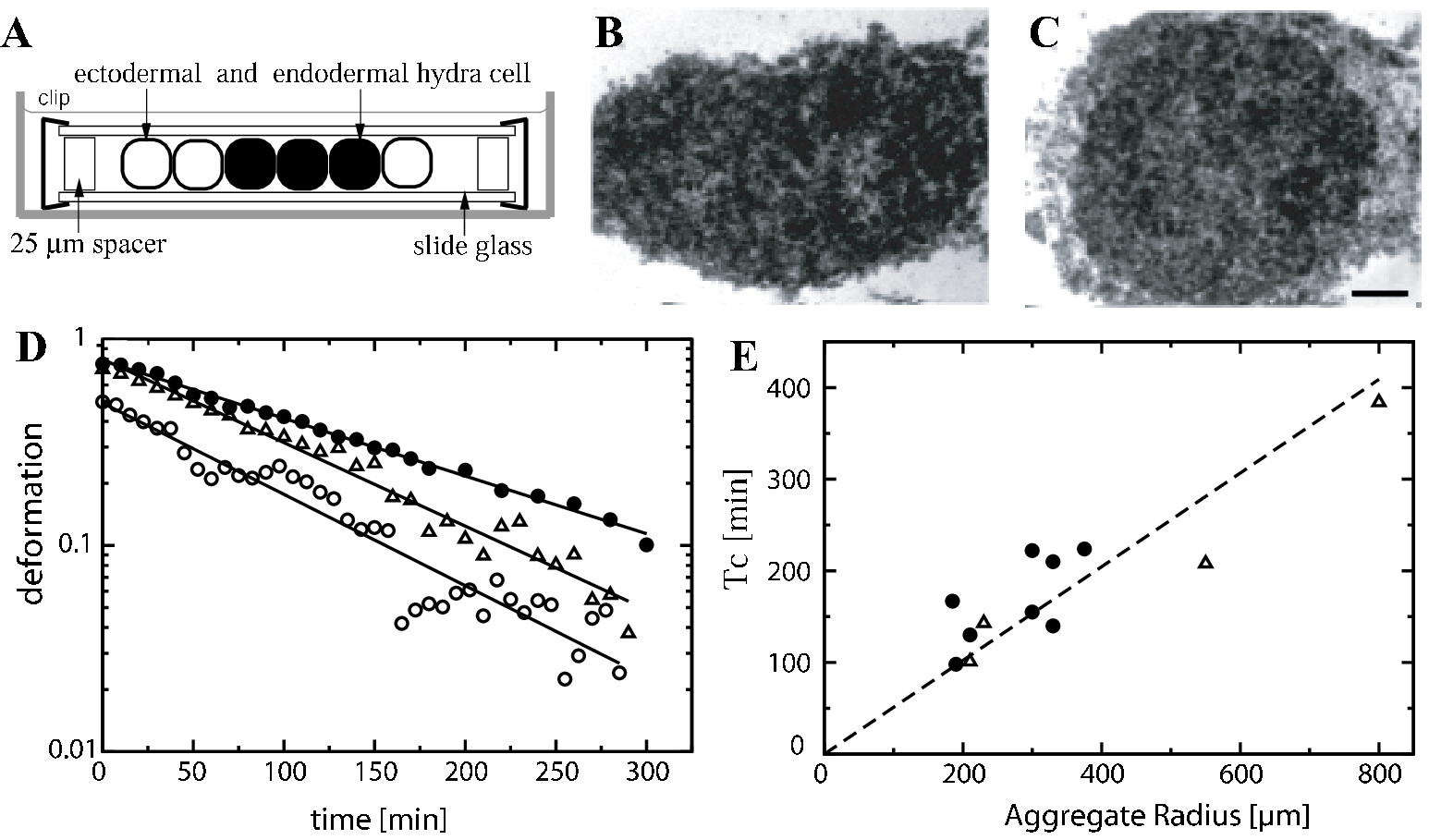}
}
\vspace{.5cm}
\caption{ 
 Shape relaxation in 2D Hydra aggregates (data
from \cite{Rieu3}). (A) 2D experimental set-up used to study
aggregate shape relaxation. (B) A pure endodermal aggregate at 0h
and (C) at 6h. Bar, 100 $\mu m$. (D) Deformation parameter $d$
(see text) as a function of time (bullets: pure endodermal
aggregate; triangles: mixed aggregate; circles: pure ectodermal
aggregate). (E) Relaxation time $T_c$ as a function of aggregate
radius (same legend as D).
}
\label{1}
\end{figure}

\begin{figure}[ht]
\centerline{
\epsfxsize 10truecm
\epsfbox%
{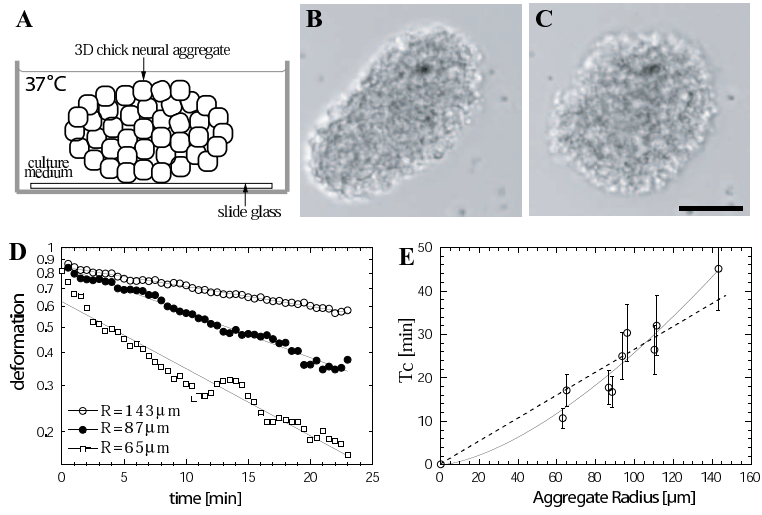}
}
\vspace{.5cm}
\caption{ 
Rounding of 3D chicken
embryonic neural retina aggregates. (A) Configuration used to
study shape relaxation: the 3D aggregates are floating near to the
bottom of glass slide in culture medium at 37$^\circ$C. (B) A
small 9-day old neural retina aggregate (R=63 $\mu m$) at 0h and
(C) at 24h. Bar, 50 $\mu m$. (D) Deformation parameter $d$ as a
function of time for three different aggregate sizes. (E)
Relaxation time $T_c$ as a function of aggregate radius. Solid
line is a power law  fit ($\sim R^{1.52}$, correlation
coefficient, 0.939), dotted line is a linear fit with with Eq.
\ref{tc3D} (correlation coefficient, 0.880).
}
\label{2}
\end{figure}

\begin{figure}[ht]
\centerline{
\epsfxsize 7truecm
\epsfbox%
{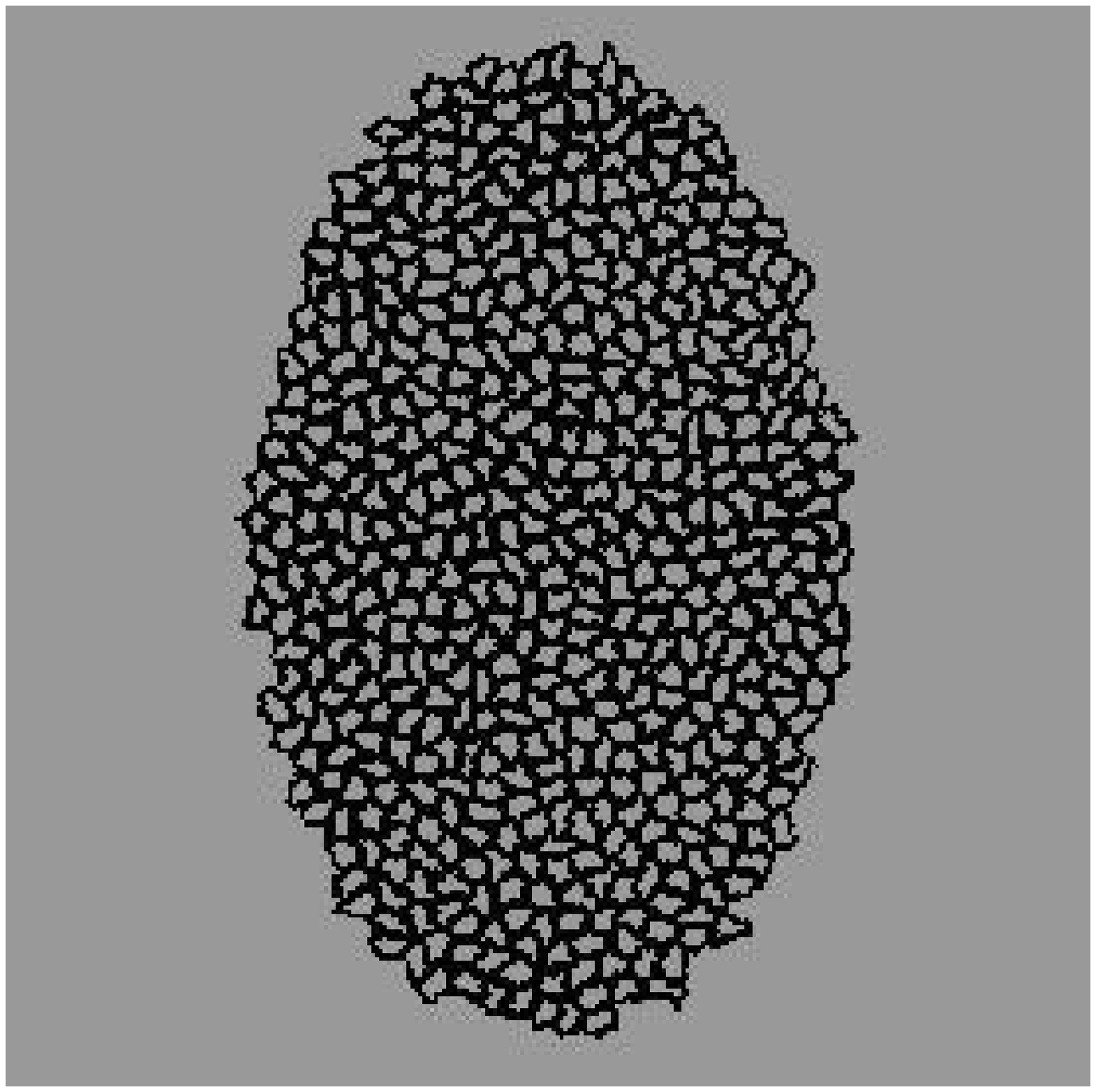}
\epsfxsize 7truecm
\epsfbox%
{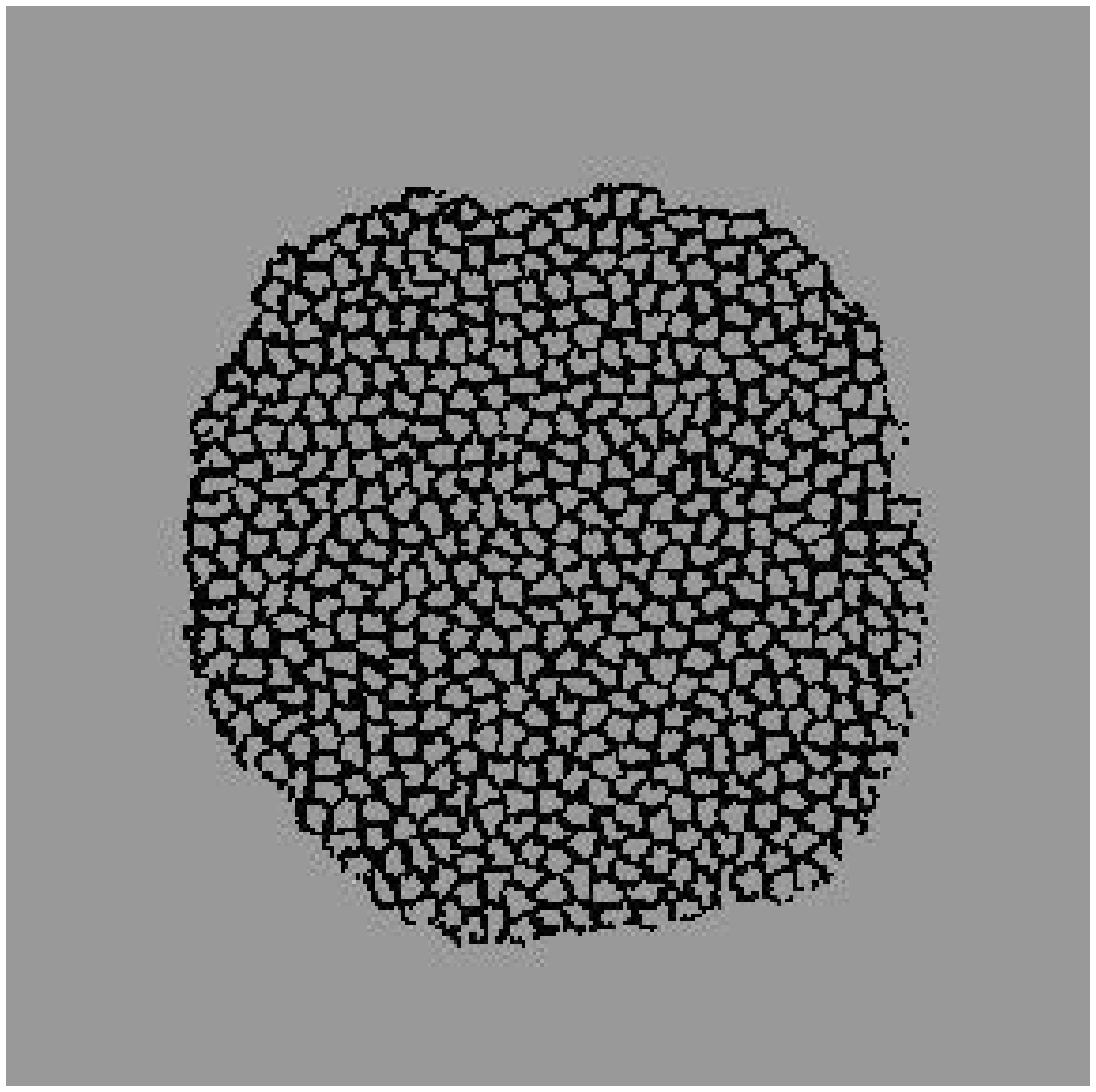}
}
\vspace{.5cm}
\caption{ 
 (A) Image of an initial
elliptical aggregate with 544 cells. (B) Final state of the
aggregate after, approximately, 10$^6$ MCS.
}
\label{3}
\end{figure}

\begin{figure}[ht]
\centerline{
\epsfxsize 10truecm
\epsfbox%
{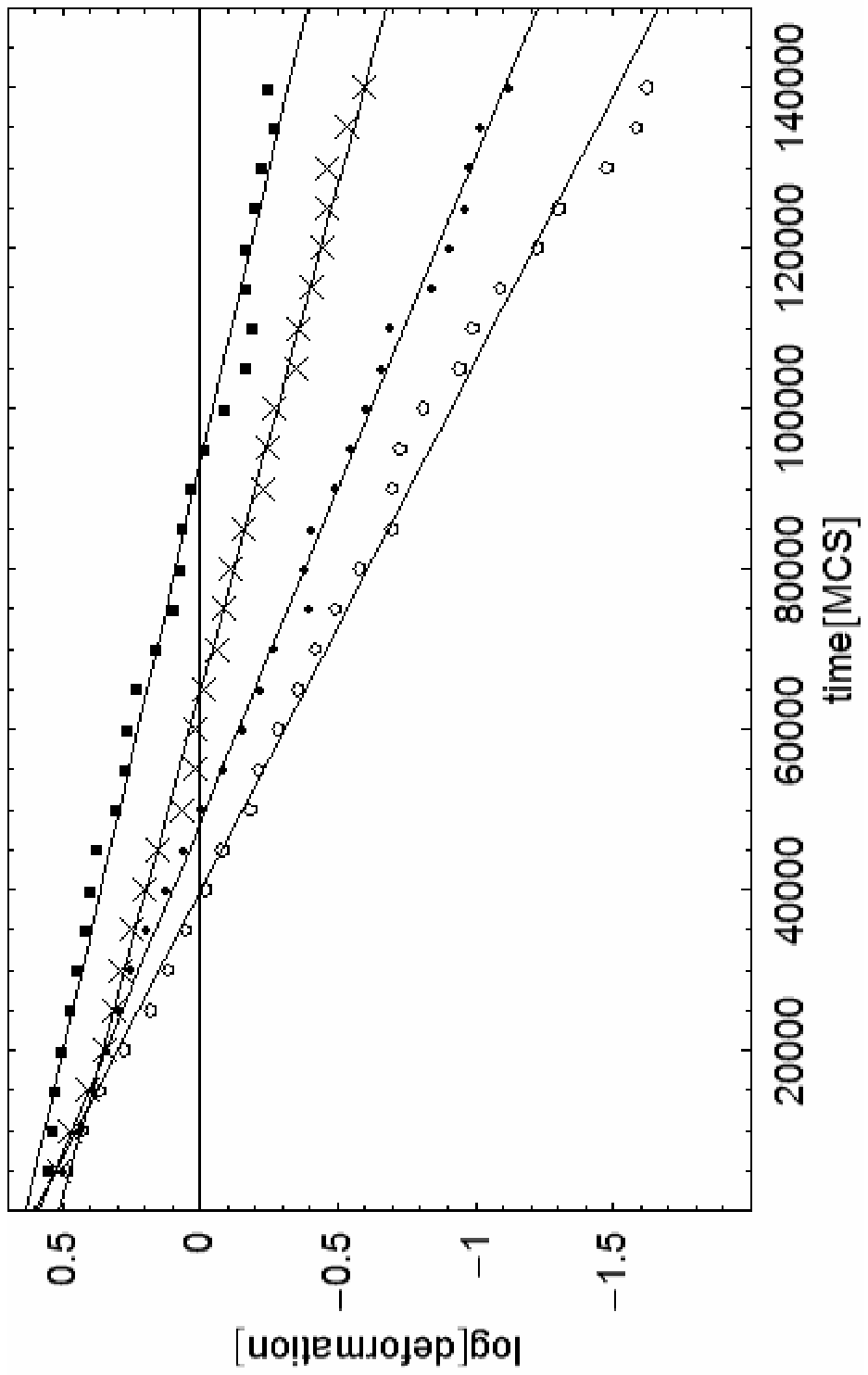}
}
\vspace{.5cm}
\caption{ 
 Log-linear plot of
deformation as a function of time for simulated aggregates with
1642 cells and surface tensions 4.5 (squares), 6.5 (crosses), 8.5
(dots), and 9.5 (circles).
}
\label{4}
\end{figure}

\begin{figure}[ht]
\centerline{
\epsfxsize 10truecm
\epsfbox%
{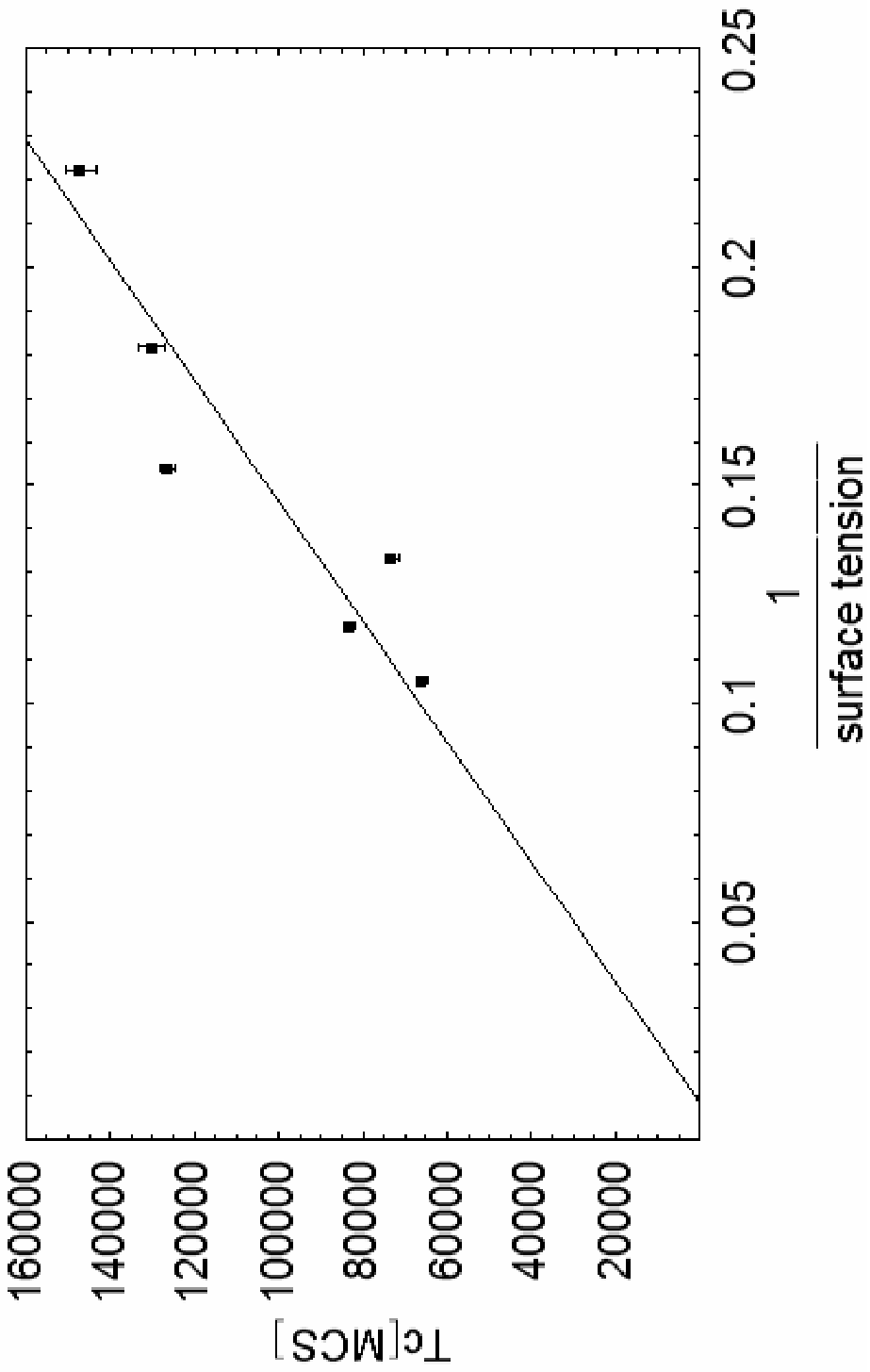}
}
\vspace{.5cm}
\caption{ 
 Plot of $T_c$ as a
function of $\sigma^{-1}$ for simulated aggregates with 1642 cells
and surface tensions 4.5, 5.5, 6.5, 7.5, 8.5, and 9.5. The
correlation coefficient of the fit is 0.85.
}
\label{5}
\end{figure}

\begin{figure}[ht]
\centerline{
\epsfxsize 10truecm
\epsfbox%
{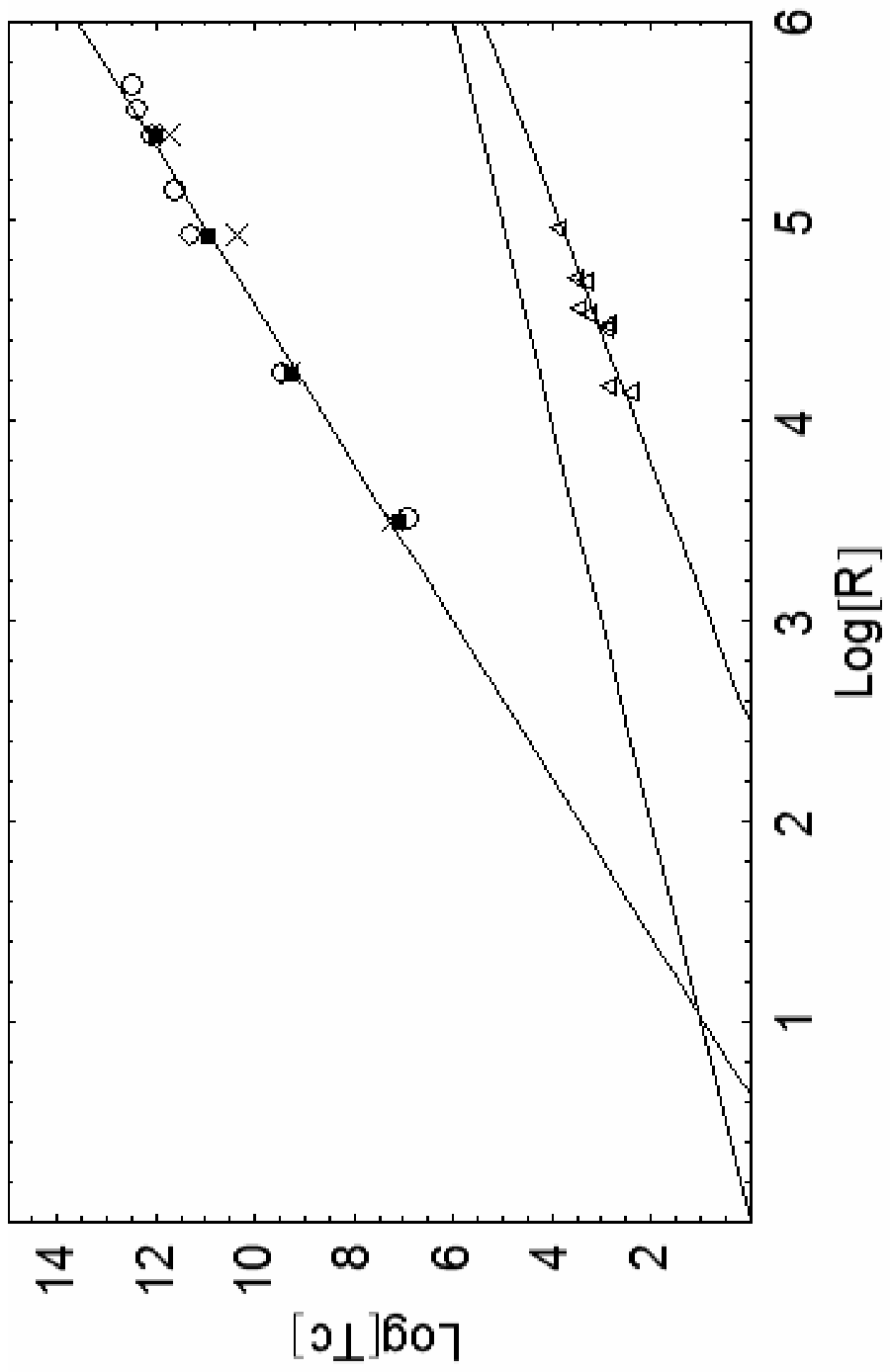}
}
\vspace{.5cm}
\caption{ 
 Log-log plot of $T_c$ vs.
$R$ for simulation and chicken data (triangles). $R$ is lattice
sites for simulation and $\mu m$ for experiments. Simulation
surface tensions are: 4.5 (circles), 5.5 (squares), 6.5 (crosses).
An identity line is shown for slope comparison.
}
\label{6}
\end{figure}

\end{document}